\documentclass[sigconf,dvipsnames]{acmart}
\usepackage[utf8]{inputenc}
\usepackage[T1]{fontenc}
\usepackage[inline]{enumitem}
\usepackage{diagbox}
\usepackage{booktabs}
\usepackage{amsmath}
\usepackage{amsfonts}
\usepackage{dashbox}
\usepackage{enumitem}
\usepackage{mathbbol}
\usepackage{mathtools}
\usepackage{hyperref}
\usepackage{url}
\usepackage{nicefrac}
\usepackage{microtype}
\usepackage{natbib}
\usepackage{multirow}
\usepackage{graphicx}
\usepackage{subcaption}
\usepackage{etaremune}
\usepackage{xcolor}
\usepackage{bbm}
\usepackage{algorithm}
\usepackage{algorithmic}
\usepackage{balance}

\newcommand{\eg}{\emph{e.g.}}

\makeatletter

\newcommand{\scriptshortto}[1][3pt]{{%
    \hbox{\rule[\scriptratio\dimexpr\fontdimen22\textfont2-.2pt\relax]
               {\scriptratio\dimexpr#1\relax}{\scriptratio\dimexpr.4pt\relax}}%
   \mkern-4mu\hbox{\let\f@size\sf@size\usefont{U}{lasy}{m}{n}\symbol{41}}}}

\makeatother

\AtBeginDocument{%
  \providecommand\BibTeX{{%
    \normalfont B\kern-0.5em{\scshape i\kern-0.25em b}\kern-0.8em\TeX}}}

\newcommand{\subtlesection}[1]{\smallskip\noindent\textbf{\emph{#1}.}}

\setcopyright{acmcopyright}
\copyrightyear{2018}
\acmYear{2018}
\acmDOI{10.1145/1122445.1122456}

\acmConference[SIGIR '21]{SIGIR 2021}{July, 2021}{Online.}
\begin{document}

\title[TREC Deep Learning Track: Reusable Test Collections in the Large Data Regime]{TREC Deep Learning Track:\\Reusable Test Collections in the Large Data Regime}


% Submissions are not to be blinded, and the author info is going to take a bunch of space.  People should check what I wrote about them to make sure it is correct! (and what you want to use) ---Ellen
\author{Nick Craswell}
\email{nickcr@microsoft.com}
\affiliation{%
   \institution{Microsoft}
%   \streetaddress{}
%   \city{}
%   \state{}
   \country{USA}
%   \postcode{}
}
\author{Bhaskar Mitra}
\email{bmitra@microsoft.com}
\affiliation{%
   \institution{Microsoft}
%   \streetaddress{}
%   \city{}
%   \state{}
   \country{Canada}
%   \postcode{}
}

\author{Emine Yilmaz}
 \email{emine.yilmaz@ucl.ac.uk}
\affiliation{%
   \institution{University College London}
   \country{UK}
}

\author{Daniel Campos}
\email{dcampos3@illinois.edu}
\affiliation{%
   \institution{University of Illinois, Urbana-Champaign}
   \country{USA}
}

\author{Ellen M. Voorhees}
\email{ellen.voorhees@nist.gov}
\affiliation{%
 \institution{National Institute of Standards and Technology}
   \country{USA}
}

\author{Ian Soboroff}
\email{ian.soboroff@nist.gov}
\affiliation{%
 \institution{National Institute of Standards and Technology}
   \country{USA}
}

\renewcommand{\shortauthors}{Craswell, Mitra, et al.}

\begin{abstract}
  The TREC Deep Learning (DL) Track studies ad hoc search in the large data regime, meaning that a large set of human-labeled training data is available. Results so far indicate that the best models with large data may be deep neural networks. This paper supports the reuse of the TREC DL test collections in three ways. First we describe the data sets in detail, documenting clearly and in one place some details that are otherwise scattered in track guidelines, overview papers and in our associated MS MARCO leaderboard pages. We intend this description to make it easy for newcomers to use the TREC DL data. Second, because there is some risk of iteration and selection bias when reusing a data set, we describe the best practices for writing a paper using TREC DL data, without overfitting. We provide some illustrative analysis. Finally we address a number of issues around the TREC DL data, including an analysis of reusability.
\end{abstract}





\maketitle

\section{Introduction}

The Text Retrieval Conference (TREC) \citep{voorhees2005trec} is an evaluation effort in the information retrieval research community that studies multiple search scenarios, called tracks. The TREC Deep Learning (DL) Track~\citep{trec2019overview, trec2020overview} studies a common search scenario where a new query comes in to search an existing corpus of text, and the goal is to produce a ranking where the most relevant search results are at the top. The distinctive aspect of the DL track is that it uses a large set of human-labeled training data. Results so far have indicated that ranking models using deep learning benefit from the large data, returning more relevant results than other types of models.

There are two tasks in TREC DL. One is passage retrieval, from a corpus of passages of text, modeling a question answering scenario where the system retrieves a short answer for the user's query. The other task is document retrieval, modeling a scenario where the user wants to see more content about their query, retrieving a ranked list of documents for the user to consider.

Each year at TREC, the two tasks each evaluate a new set of test queries, producing two reusable test collections. If a research paper is studying document retrieval and implements a baseline ranker and a new ranker, these two rankers can be compared on the reusable test sets. The standard approach is to report the results on each test set separately and use statistical tests to see if the new approach is better than the baseline approach.

This paper describes the reusable test collections built as part of the DL Track, bringing together and summarizing descriptions that previously existed only in DL Track guidelines, the overview papers and in our Github repositories. Then, since it is possible to overuse a test data set, we describe the best practices for using the test sets. We present a case study of making decisions using the dev set, despite it having a different labeling scheme from the test sets, showing it is predictive of test set performance without overfitting to test data. Since we focus on reuse, we also present a reusability analysis of the test sets, indicating that judgments are sufficiently complete to evaluate a new ranker without any need for additional relevance judgments. We finish by pointing out some limitations of the current setup, for example it focuses on only on our two tasks, only one language.

\begin{figure*}
  \center
  \begin{subfigure}{.39\textwidth}
    \includegraphics[width=\textwidth]{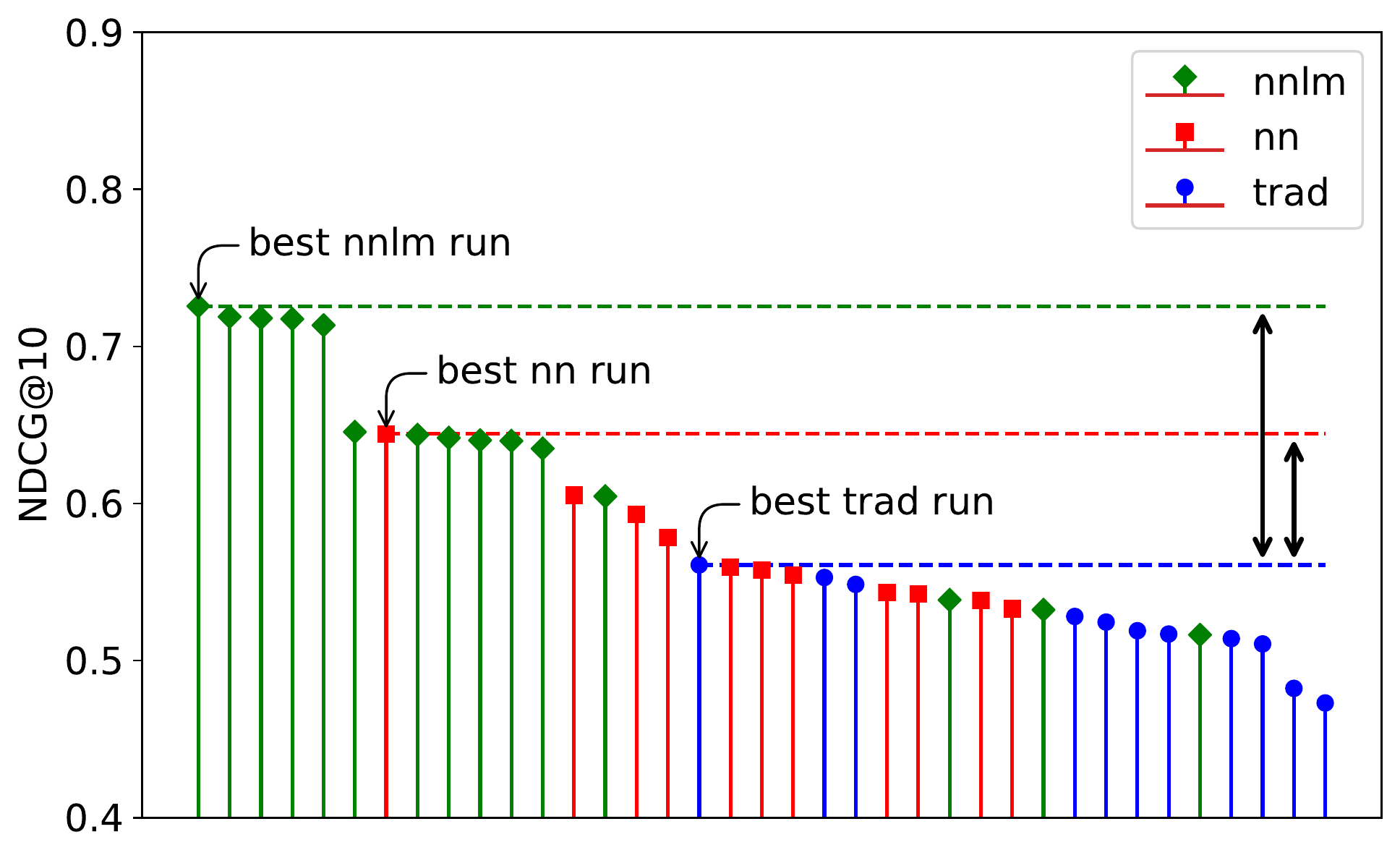}
    \caption{Document retrieval task (2019)}
    \label{fig:model-task-docs-stem-by-model-type-2019}
  \end{subfigure}
  %\hfill
  \begin{subfigure}{.39\textwidth}
    \includegraphics[width=\textwidth]{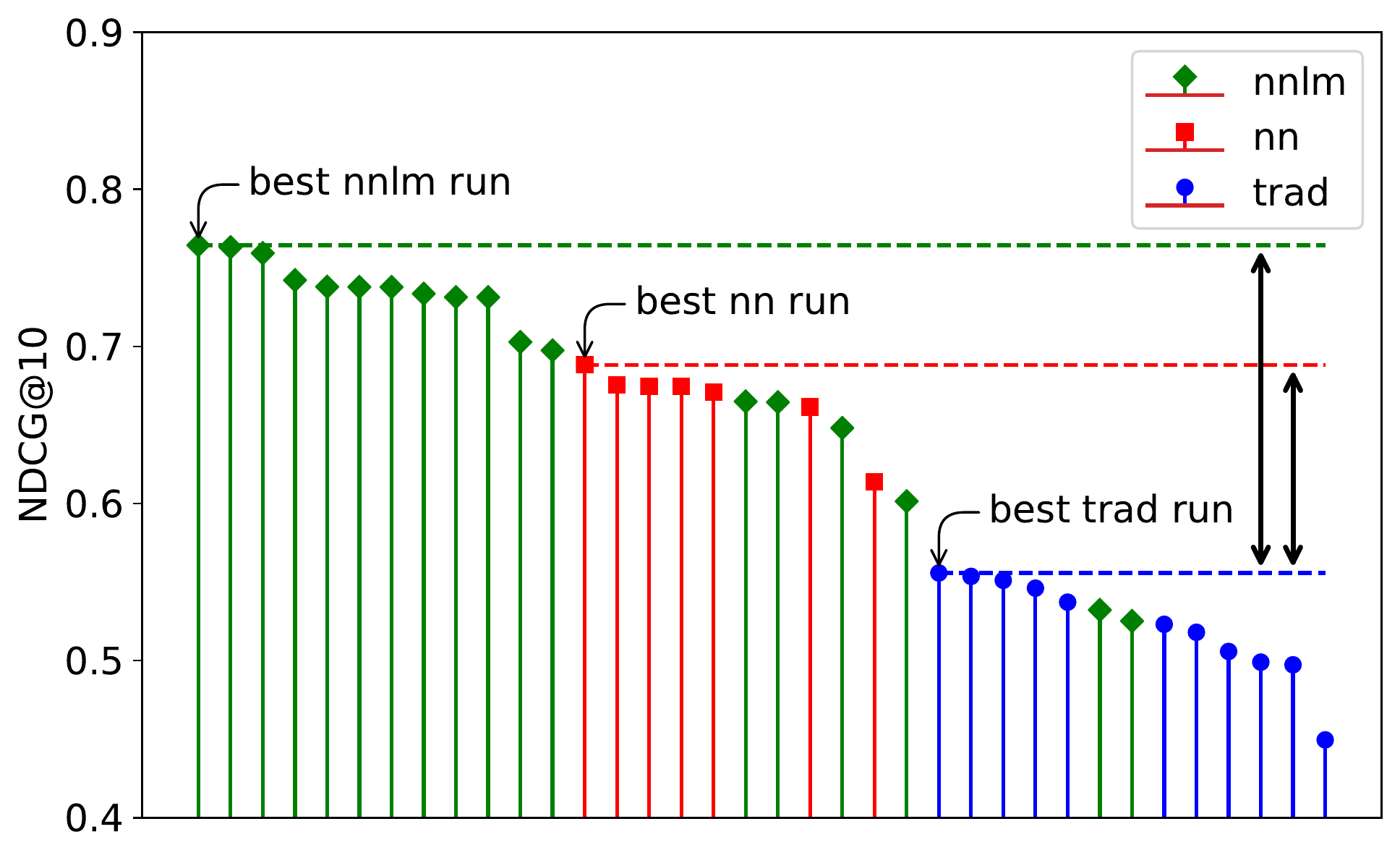}
    \caption{Passage retrieval task (2019)}
    \label{fig:model-task-passages-stem-by-model-type-2019}
  \end{subfigure}
  \center
  \begin{subfigure}{.39\textwidth}
    \includegraphics[width=\textwidth]{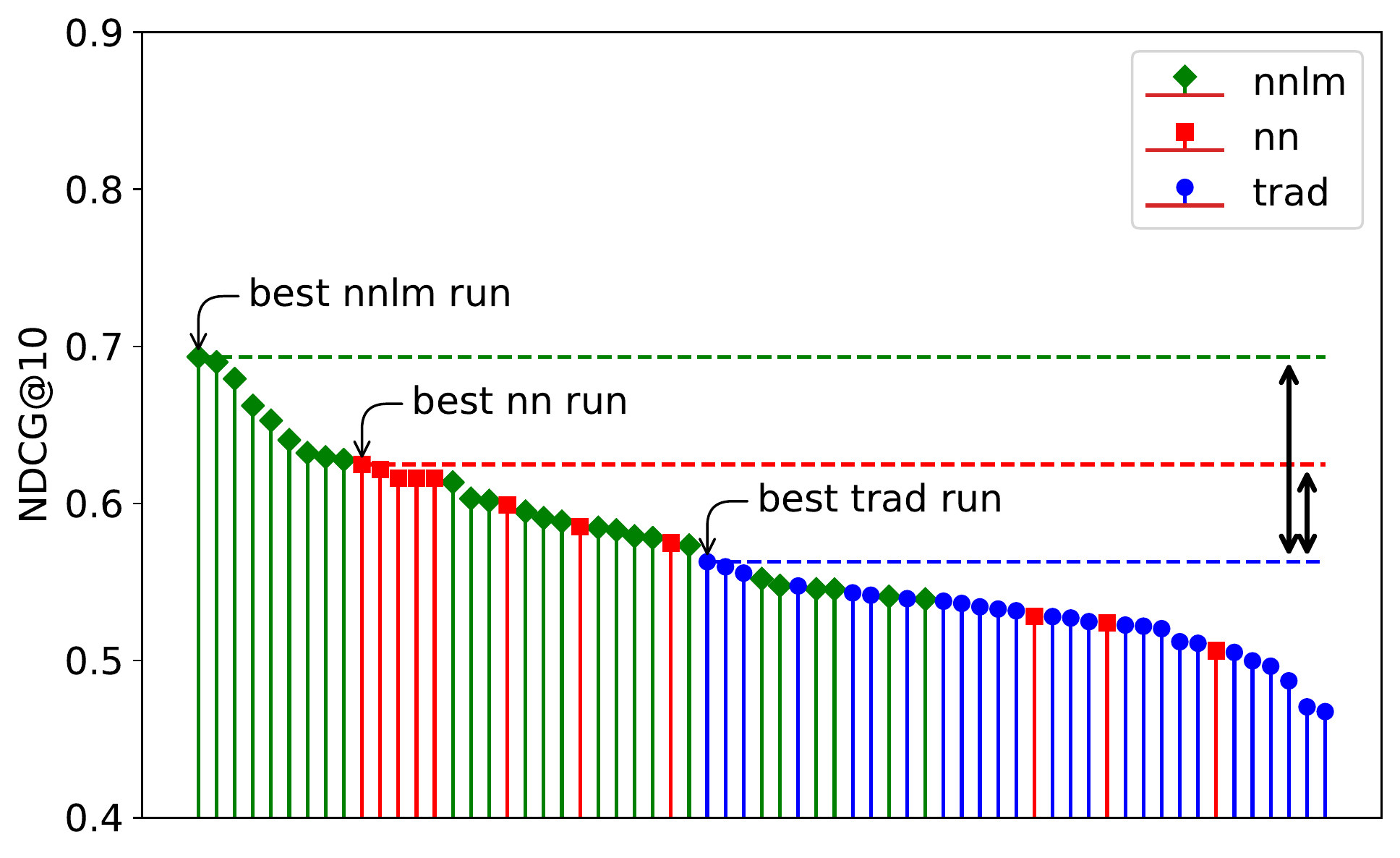}
    \caption{Document retrieval task (2020)}
    \label{fig:model-task-docs-stem-by-model-type-2020}
  \end{subfigure}
  %\hfill
  \begin{subfigure}{.39\textwidth}
    \includegraphics[width=\textwidth]{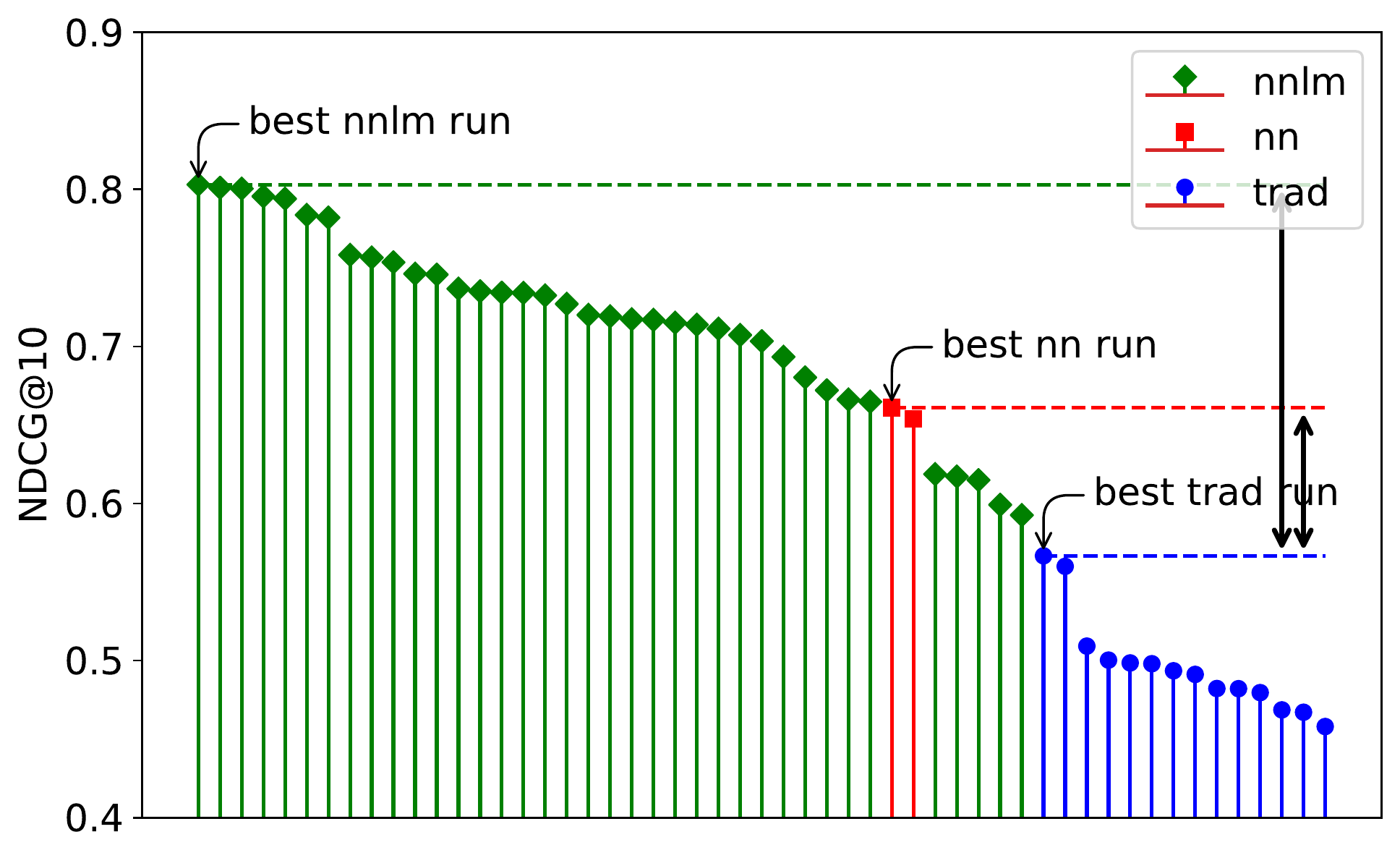}
    \caption{Passage retrieval task (2020)}
    \label{fig:model-task-passages-stem-by-model-type-2020}
  \end{subfigure}
  \caption{NDCG@10 results for TREC submissions, broken down by run type. BERT-style ``nnlm'' runs performed best in both tasks in both years, with non-BERT ``nn'' runs and non-neural ``trad'' runs having relatively lower performance.}
  \label{fig:model-stem-by-model-type}
\end{figure*}

%\todo{Do we need a related work section?}
% Nick: We decided to just cite related work throughout, since the main related work is TREC

\section{Task description}
The Deep Learning track has two tasks: Document retrieval and passage retrieval.

\subtlesection{Document retrieval task}
The first task focuses on document retrieval, with two subtasks:
\begin{enumerate*}[label=(\roman*)]
    \item Full retrieval and
    \item Top-$100$ reranking.
\end{enumerate*}
In the full retrieval subtask, the retrieval system is expected to rank documents based on their relevance to the query, where documents can be retrieved from the full document collection provided.
This subtask models the end-to-end retrieval scenario.
In the reranking subtask, the benchmark provides an initial ranking of $100$ documents, giving all ranking methods the same starting point.
This is a common scenario in many real-world retrieval systems that employ a telescoping architecture~\citep{matveeva2006high, wang2011cascade}.
The reranking subtask allows participants to focus on learning an effective relevance estimator, without the need for implementing an end-to-end retrieval system.
It also makes different reranking approaches more comparable, because they all rerank the same set of 100 candidates.
The initial top-$100$ rankings were retrieved using Indri~\citep{strohman2005indri} on the full corpus with Krovetz stemming and stopwords eliminated.

\subtlesection{Passage retrieval task}
Similar to the document retrieval task, the passage retrieval task includes
\begin{enumerate*}[label=(\roman*)]
    \item Full retrieval and
    \item Top-$1000$ reranking.
\end{enumerate*}
In the full retrieval subtask, given a query, the retrieval system is expected to retrieve a ranked list of passages from the full collection based on their estimated likelihood of containing an answer to the question.
In the top-$1000$ reranking subtask, $1000$ passages per query are provided, giving all methods the same starting point.
The sets of 1000 were generated based on BM25 retrieval with no stemming as applied to the full collection.
All ranking models are expected to rerank the 1000 passages based on their estimated likelihood of containing an answer to the query.
In this subtask, we can compare different reranking methods based on the same initial set of $1000$ candidates, with the same rationale as described for the document reranking subtask.

\section{Test Collection Data}
\label{sec:data}

A test collection comprises a corpus, test queries and test relevance judgments \cite{voorhees2005trec}. We have a document corpus and a passage corpus, each with two test sets so far and a large training set, as summarized in Table~\ref{tbl:data}. We now describe the data in more detail.

%Corpus. Queries. Training set. Dev set. TREC test sets. All the statistics.
\subsection{Training Data Sets}
\label{sec:data-train}
Both tasks have large training sets based on human relevance assessments, derived from MS MARCO~\citep{bajaj2016ms}.
These are sparse, with no negative labels and often only one positive label per query, analogous to some real-world training data such as click logs.
In the case of passage retrieval, the positive label indicates that the passage contains an answer to a query. 
In the case of document retrieval, we transferred the passage-level label to the corresponding source document that contained the passage.
We do this under the assumption that a document with a relevant passage is a relevant document, although we note that our document snapshot was generated at a different time from the passage data set, so there can be some mismatch.
Despite this, machine learning models trained with these labels seem to benefit from using the labels, when evaluated using NIST's non-sparse, non-transferred labels.
This suggests the transferred document labels are meaningful for our TREC task.

\subsection{Test Sets}
\label{sec:data-test}
The test collections were constructed using shallow pooling across all runs submitted to the respective task, with additional documents selected to be judged via the HiCAL classifier~\cite{HiCAL}. The test set of 200 queries was the same for both tasks.  A subset of those queries were selected to be judged at NIST based on effectiveness as scored using the MARCO sparse judgments---queries with a median MRR across document retrieval submissions of greater than 0.5 or of 0.0 were excluded from the evaluation set.  Some evaluation-set candidate queries were later eliminated because their number of relevant documents fell outside the range that would create a robust test collection.  In the end, the evaluation set contained 43 queries for both tasks in 2019, though they were different sets of queries. The same judgment process was used in 2020, which resulted in evaluation sets of 45 and 54 queries for the document and passage retrieval tasks, respectively.

For the document retrieval task, judgments were on a four-point scale of \textbf{Irrelevant}, \textbf{Relevant}, \textbf{Highly Relevant}, and \textbf{Perfectly Relevant}.  For measures that use binary judgments, all but Irrelevant are counted as relevant. Passage retrieval judgments were also collected on a four-point scale: \textbf{Irrelevant}, \textbf{Related}, \textbf{Highly Relevant}, and \textbf{Perfectly Relevant}.  In this case, Related means the passage is on-topic, but does not actually answer the question; hence only Highly and Perfectly Relevant are treated as relevant for binary measures.

When reusing the test sets, there can be a significant problem if the new ranker retrieves documents that are relevant but unjudged. This makes it difficult to correctly estimate the quality of the new model. Since reusability is a focus in this paper, we provide some more in-depth reusability analysis in Section~\ref{sec:reusability analysis}.

\subsection{ORCAS}
\label{sec:data-orcas}
The 2020 edition of the track also released a large scale click data set for the document retrieval task.
The ORCAS data~\citep{craswell2020orcas} is constructed from the logs of a major search engine.
The data can be used in a variety of ways, for example as additional training data (almost 50 times larger than the main training set) or as a document field in addition to title, URL and body text fields available in the original training data.
However, we do not describe the ORCAS data in details here and instead point the interested reader to the ORCAS website\footnote{https://microsoft.github.io/msmarco/ORCAS}.

\subsection{TREC runs}

The TREC Deep Learning Track had 15 and 25 participating groups---with a total of 75 and 123 runs submitted across both tasks---in 2019 and 2020, respectively. Since runs are blind submissions, that are finalized before any relevance labels are available, they provide a comparison of ranking approaches with no chance of overfitting to the test judgments.

Based on submission surveys with each run, we divided the runs into three categories:
\begin{enumerate*}
    \item \textbf{nnlm:} if the run employs large scale pre-trained neural language models, such as BERT \citep{devlin2018bert} or XLNet \citep{yang2019xlnet}
    \item \textbf{nn:} if the run employs some form of neural network based approach---\eg, Duet \citep{mitra2017learning, mitra2019updated} or using word embeddings \citep{joulin2016bag}---but does not fall into the ``nnlm'' category
    \item \textbf{trad:} if the run exclusively uses traditional IR methods like BM25 \citep{robertson2009probabilistic} and RM3 \citep{abdul2004umass}.
\end{enumerate*}

Overall results (Figure~\ref{fig:model-stem-by-model-type}) in both tasks in both years were that the best ``nnlm'' runs outperformed other types of run. This effect was more pronounced in passage retrieval, since the chances of vocabulary mismatch between query and result is greater if the search result is a shorter text. The track overview papers \cite{trec2019overview,trec2020overview} present a number of more detailed breakdowns, for example by subtask. We note that NIST makes the runs available, so they become part of the DL Track resources. Past TREC runs could be studied as baselines.

\subsection{Reusing the Test Collections}
\label{sec:data-reuse}
\label{sec:using-the-test-collections}
Table \ref{tbl:data} provides descriptive statistics for the data set.
More details about the data sets---including directions for download---is available on the TREC 2020 Deep Learning Track website\footnote{\url{https://microsoft.github.io/msmarco/TREC-Deep-Learning}}.
Interested readers are also encouraged to refer to \citet{bajaj2016ms} for details on the original MS MARCO data set.

\begin{table}[]
\centering
\caption{Summary of statistics on TREC 2020 Deep Learning Track data sets.}
\begin{tabular}{@{}lrr@{}}
\toprule
~                     & \multicolumn{1}{c}{Document task} & \multicolumn{1}{c}{Passage task} \\ 
Data                  & \multicolumn{1}{c}{Number of records} & \multicolumn{1}{c}{Number of records} \\ 
\midrule
Corpus                & $3,213,835$    & $8,841,823$       \\
\addlinespace
Train queries         & $367,013$      & $502,939$         \\
% Train candidates      & $36,701,116$   & $478,016,942$               \\
Train qrels           & $384,597$      & $532,761$         \\
\addlinespace
Dev queries           & $5,193$        & $6,980$          \\
% Dev candidates        & $519,300$      & $6,668,967$          \\
Dev qrels             & $5,478$        & $7,437$          \\
\addlinespace
2019 TREC queries    & $200 \rightarrow 43$          & $200 \rightarrow 43$             \\
% Validation candidates & $20,000$       & $59,273$          \\
2019 TREC qrels      & $16,258$       & $9,260$          \\
\addlinespace
2020 TREC queries          & $200 \rightarrow 45$          & $200 \rightarrow 54$             \\
2020 TREC qrels            & $9,098$    &  $11,386 $        \\
%Test candidates       & $20,000$       & $190,699$         \\ 
\bottomrule
\end{tabular}
\label{tbl:data}
\end{table}

Broadly speaking, the data set contains four separate sets: train, dev, TREC 2019 test, and TREC 2020 test.
The relevance judgments in train and dev sets are binary, while for TREC 2019 and 2020 test sets they are on a four-point scale as described in Section~\ref{sec:data-test}.
A typical use of this data set would involve traininig a model on the large training corpus, using the dev set (or a subset of it) to make decisions about hyperparameter choices and early stopping, and evaluating and reporting the model performance on the TREC 2019 and 2020 test sets.
However, multiple evaluation of different model variants on the TREC test sets can lead to problematic outcomes, such as, overfitting on these query sets and false positive results on model performance.
To avoid this, we strongly recommend not to iterate over the TREC sets.
Instead, a more reasonable experiment protocol would be to use the dev set to validate and iterate on new architecture and other proposed changes.
Only after the final model has been selected for publication, should they be evaluated against the TREC test sets to generate the final numbers that can be reported in a publication or in other forums.
Section~\ref{sec:validity} provides evidence that supports the validity of conclusions reached via such an experimentation protocol.

%How to resuse. Describe the code here too? Train your model on the training set. Select what rankers go in your paper using the dev set. Finally, run the test sets and submit your paper. Do not iterate.

%\todo{Fullrank vs rerank. One-stage vs multi-stage. 'trad' vs 'nn' vs 'nnlm' and including the concept of 'pure nnlm' has no trad stuff.}

\subsection{Code}
\label{sec:data-code}
To make it easier to work with this benchmark, we open-source a PyTorch~\citep{paszke2017automatic} implementation\footnote{https://github.com/bmitra-msft/TREC-Deep-Learning-Quick-Start} of the Conformer-Kernel model~\citep{mitra2020conformer, mitra2020conformer2}.
The code automates the download of all the required data files for the document retrieval task.
It also implements the training and evaluation of a relatively efficient deep neural baseline on this benchmark, under both the rerank and the fullrank settings.
This code is provided as-is primarily for the convenience of those working with this data set for the first time, but it is not compulsory to use this codebase in the context of this benchmark.

\begin{figure}
    \centering
    \includegraphics[width=\columnwidth]{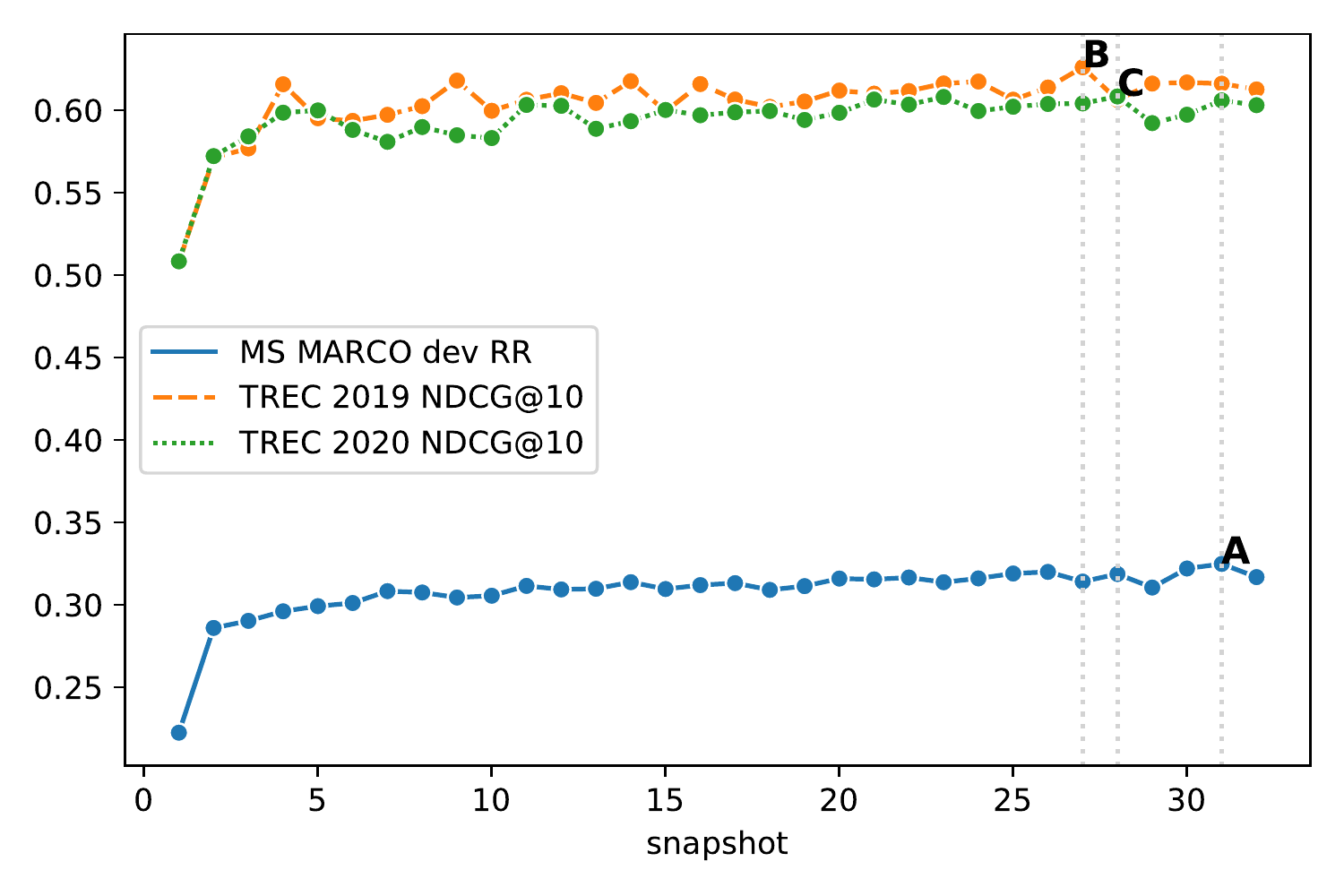}
    \caption{Training curves. Using 32 snapshots of the training process, we have 32 candidate rankers. Rankers A, B and C are optimal on dev RR, 2019 NDCG@10 and 2020 NDCG@10 respectively. The correct ranker to select is A, then report results of ranker A on the 2019 and 2020 test queries.}
    \label{fig:training_curves}
\end{figure}

\begin{figure}
    \centering
    \includegraphics[width=\columnwidth]{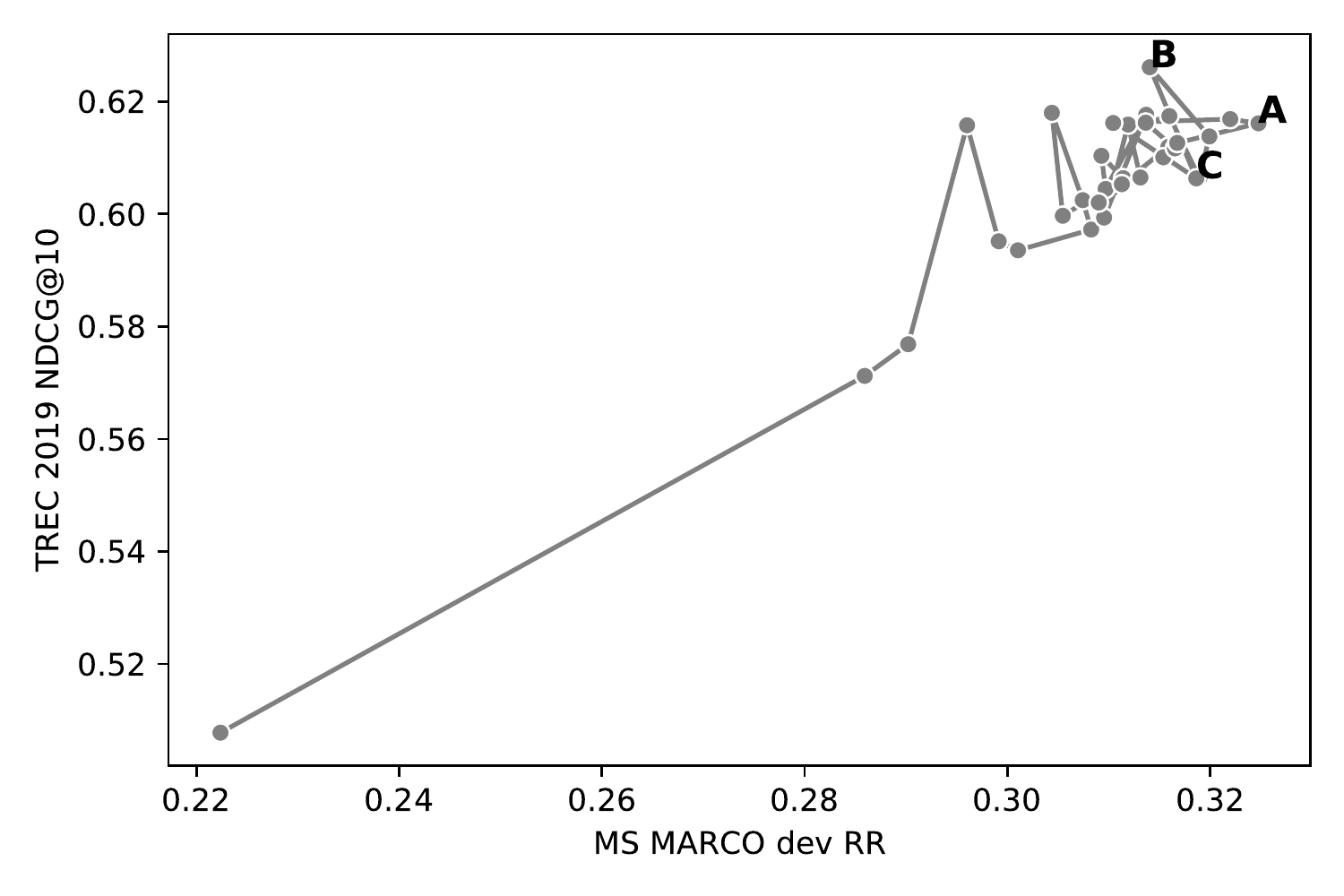}
    \includegraphics[width=\columnwidth]{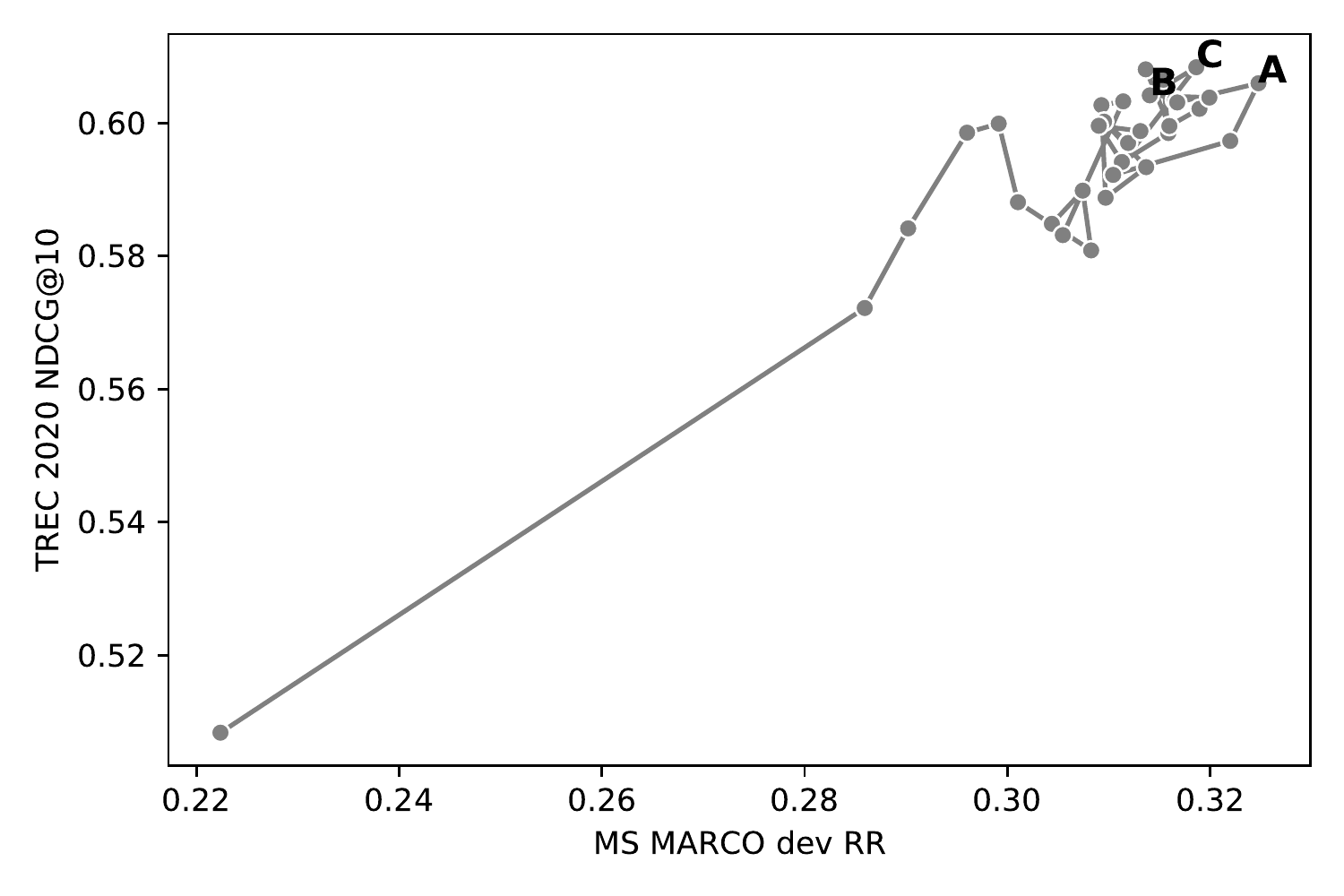}
    \includegraphics[width=\columnwidth]{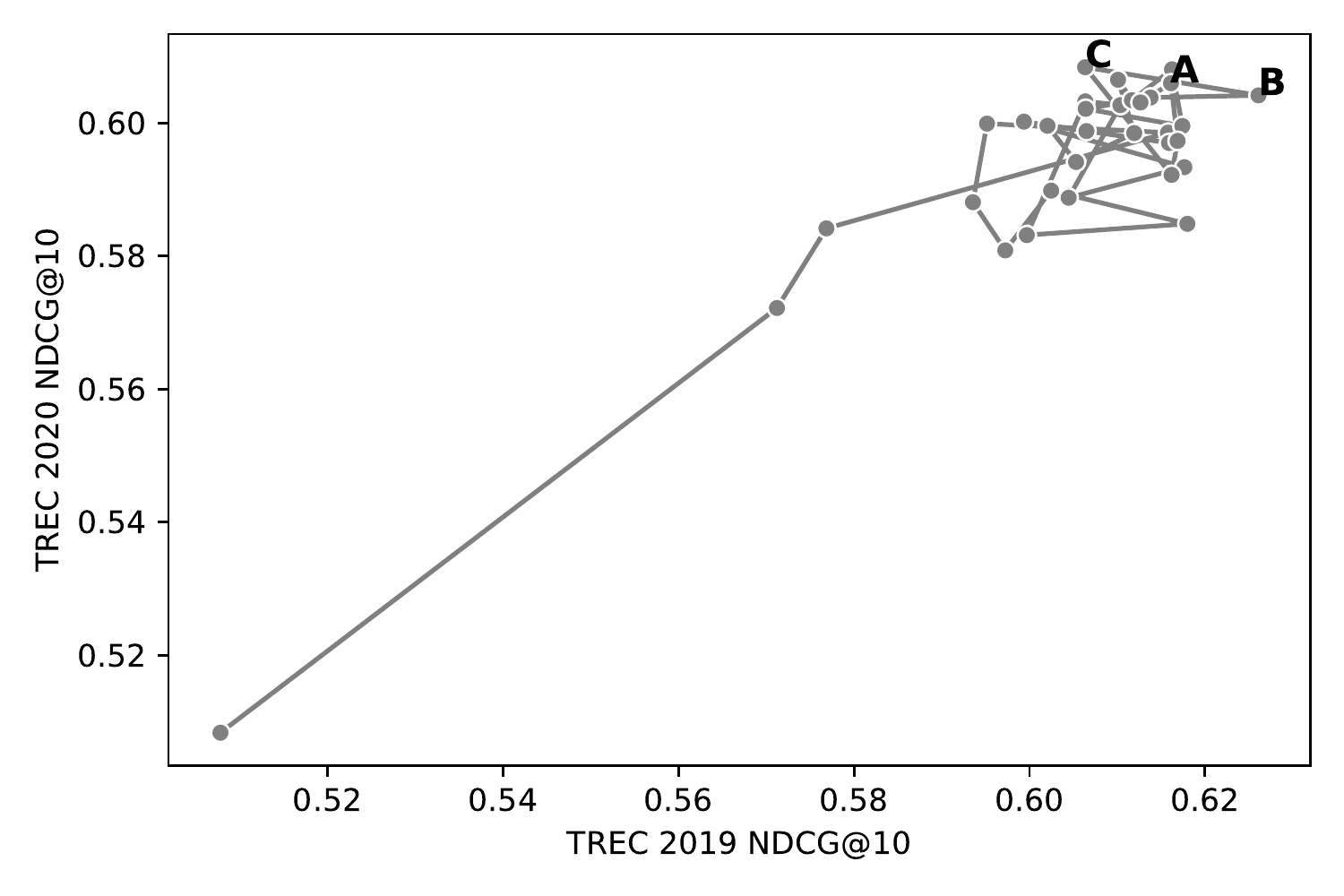}
    \caption{Results vary, even in a single trail of training a model. Select your rankers using the dev set, which in this case is ranker A, and report the results on TREC 2019 and TREC 2020. Selecting your ranker based on the TREC sets (B and C) is not acceptable.}
    \label{fig:training_scatter}
\end{figure}

\section{Validity of Results}
\label{sec:validity}

% https://www.mitpressjournals.org/doi/pdf/10.1162/neco.1997.9.5.1143
% Rich: Figure 2, Figure 4

The easy reusability of TREC test collections is extremely important, but can also cause problems. The important advantage of reusability is that we can test a new hypothesis on several years of TREC data without waiting years for new iterations for TREC. If the same ranking approach gives positive results on multiple test sets, we are more confident that it is a real improvement, rather than a lucky one that was positive on a single test.

The negative side of reusability is related to selection bias, harming the validity of results. In one scenario, the researcher tries their hypothesis on reusable test collections, the results are negative and their paper is rejected. This means that published papers are biased towards positive results. Some positive results will be false positives, where the researcher was lucky. In another scenario, the researcher gets a negative result, so they try another configuration or variant of their hypothesis. With enough iteration and multiple testing, they get results good enough to publish in a paper and the paper is accepted. With multiple testing, the chances of getting lucky are increased, and the chances of their result being correct are reduced. Ideally we should make no decisions using the test sets, much less a series of iterations. We should make decisions on the dev set.

To illustrate this, we present a case study. It is a singe training run of the Conformer-Kernel approach described in Section~\ref{sec:using-the-test-collections}. We use default parameters, but take a checkpoint of the ranker training every 1024 samples instead of 4096, to get 32 checkpoints during training (Figure~\ref{fig:training_curves}). Each checkpoint is a ranker that could be presented in a paper. The rankers do not differ in an interesting way, since they all come from the same training setup, and the only difference is the stopping point of the training.

We recommend to choose the ranker with the best dev set performance, which here is the 31st checkpoint. That ranker is indicated by a vertical dotted line and letter A in the figure. Since we used the dev to make a decision, the dev set numbers are not a valid test to report in the paper. Whenever we have several options and choose the best-performing option on a test set, that test set is no longer valid. The valid numbers to report are the TREC 2019 and TREC 2020 NDCG@10 results for ranker A, which can be read by following the dotted line from A to the two other curves. By the same token we should not publish results of the 27th or 28th iteration, which are rankers B and C, unless we want to choose results on one TREC test set and report results on the other set.

Figure~\ref{fig:training_scatter} shows the same 32 rankers, but focusing on the variability and correlation of the different metrics. The metrics are largely correlated, but between checkpoints there is also random variation. So even within one trial of ranker training, if we allow cherrypicking of the best checkpoint, we can get a meaningless positive result. We note that ranker A is near-optimal on both the TREC collections, so the dev set is working well if our goal is to choose a good checkpoint. We also note that rankers B and C are quite different from each other on the TREC 2019 data, and the difference is statistically significant. There is no sensible procedure that would lead to the comparison of rankers B and C in a paper. Instead, this illustrates that there is enough random variation in a single training run that we can find a completely meaningless `significant' ($p=0.00837$) difference between two rankers.

The general principle is to avoid making decisions on the test set. Throughout the process of experimentation for a paper it is better to make multiple use of the dev set, and only try methods on the test set when they are finalized.

Since making decisions on the same test set can lead to unreliable conclusions, another approach that could be possible is to divide the dev set into two or three parts. One set could be used for choosing the checkpoint as in our case study. Another set could be used to compare which neural network architecture is better. Becuase we split the data, the architecture question can be asked independently of the checkpoint decision. In the case of a third set it would even be possible to publish dev set results, since we have a third set that was never used for choosing between checkpoints or choosing architectures. To this end, we plan to publish some standard splits of the dev set, to enable publication of dev results on the same held out test set.

Rather than comparing a particular machine learned ranker to a baseline ranker, our goal should be to compare an overall machine learning approach for ranking to a baseline approach. The significant variation we see in a single training run (Figure~\ref{fig:training_scatter}) reminds us that we can better understand the performance of a learning approach by running multiple training trials with different random seeds. Considering the performance across trials increases the chances of finding a true difference between machine learning approaches, by decreasing the chances that results are dominated by a lucky or unlucky random seed. \citet{ganjisaffar2011bagging} provide one demonstration of this approach being applied in information retrieval.

Overall, the best ranker on a test set is lucky \citep{carterette2015best}. Even selecting the best baseline from a set of simple ``trad'' rankers can seem like overfitting, since the best ranker varies depending on the test data \citep{zobel1998exploring}. Instead we should strive to identify ranking approaches that work well across a variety of test sets, without making any design decisions using these particular test sets. This maximizes the chances that the models will perform well when deployed, on an independent future sample of test queries.

% Sensitivity. Dev set is always SS between A-B-C, for example. This is good, people can get some early guidance on what's working. When they try TREC, what if the gains are non-SS. Guidance on effect size or power, to make sure the gains are big enough on dev to appear on TREC.

% Also: Do multiple trials.

\begin{figure*}
    \centering
    \includegraphics[width=3in,height=1.75in]{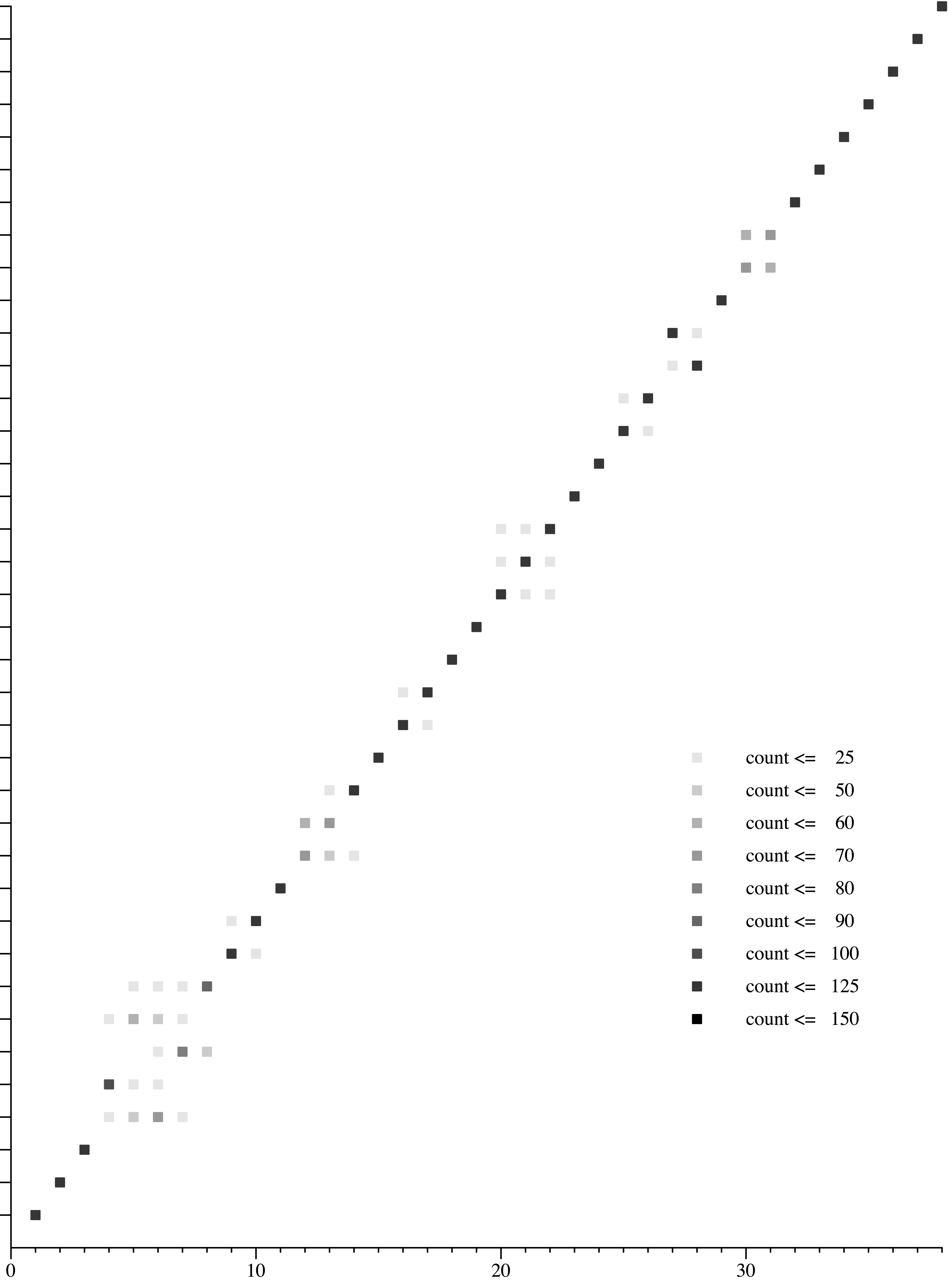}\hfill
    \includegraphics[width=3in,height=1.75in]{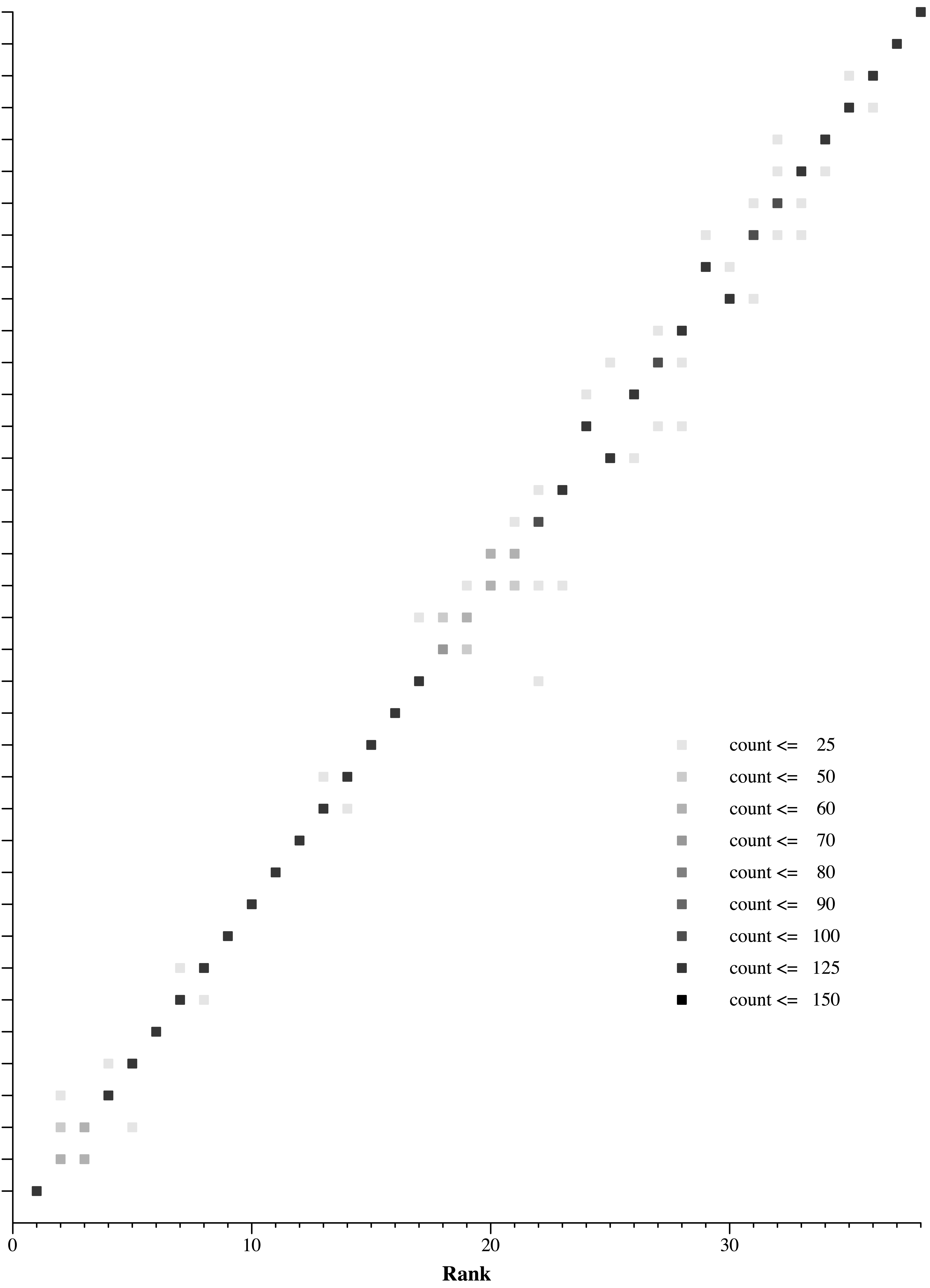}
    
    \hspace {.5in} a) Document retrieval task, MAP\hfill
    b) Document retrieval task, Precision@10\hspace{.4in}
    \caption{Heatmap of ranks obtained by a run over 120 trials for the 2019 document task when using MAP (left) or Precision@10 (right) as the evaluation measure. The darker a plotted point, the more times the run was ranked at that position.}
    \label{runranks.fig}
\end{figure*}

\section{Reusability analysis}
\label{sec:reusability analysis}

One of the goals of the track is to build general-purpose, reusable test collections at acceptable cost. In this context, general-purpose means a collection reliably ranks runs for a wide spectrum of evaluation measures, including recall-focused measures. Reusable means that runs that did not participate in the collection building process can be reliably ranked by the collection. In this section we support the claim that the DL track collections are reusable.

Leave-Out-Uniques (LOU) tests~\cite{IRPooling,zob:sigir98} are a traditional way of analyzing the reusability of a collection. In these tests, the relevant documents retrieved by only one participating team are removed from the qrels files and all runs are then evaluated using the reduced qrels. The reduced qrels are the qrels that would have resulted had the team not participated in the collection building process, and thus their submitted runs represent new runs with respect to the reduced qrels. New runs ranking essentially the same in both the original and reduced collections supports a claim that the collection is reusable.

However, a standard LOU test does not work for the collections built in the DL track because of the way the judgment sets were constructed. We used the HiCAL system~\cite{HiCAL} to select the set of documents to be judged for each query once depth-10 pools were judged. HiCAL uses the current set of judgments to build a relevance model and then selects the unjudged document most likely to be relevant as the next document to judge. HiCAL does not depend on runs as the source of documents, so the concept of uniquely retrieved relevant documents no longer applies.

A given team's unique relevant documents can be removed from the depth-10 pools in the first stage, but then the HiCAL process must be activated as it may select the removed documents to be judged. Since the HiCAL process is not deterministic (ties are broken randomly) and depends on the particular set of documents seen so far, the HiCAL process must be simulated multiple times using the original qrels' judgments for a fair test.

The simulations proceeded as follows, where the entire process was performed separately for the each DL track task. The original depth-10 pools were fed to HiCAL for each of ten trials, where each trial used a separate initial seed for the random number generator. Within each trial, we tracked the documents encountered by HiCAL, creating a trace of the first 2500 documents encountered per query. Any unjudged documents encountered by HiCAL were treated as not relevant. We created a qrels file from each trace by taking a prefix of the trace of length equal to the number of documents judged in the original qrels per query. This resulted in 10 qrels files that could have been the result of the official judgment process of the track (modulo the unjudged documents would have been judged). 

To obtain the effect of the LOU test, for each team in turn we omit that team's uniquely retrieved relevant documents from the depth-10 pools. This pool is fed to the HiCAL process for each of ten trials where the random number seed for a given trial is the same as in the all-teams simulation.  As before, we create a trace of the documents that were encountered by HiCAL, and create a qrels file by taking a prefix of the trace of length equal to the number of documents judged in the official qrels.  All runs are evaluated using this trial's qrels, and the system ranking induced by it for a given evaluation measure is compared to the ranking induced by the official qrels.

Figure~\ref{runranks.fig} shows the results of this modified LOU test for the TREC 2019 document task (the passage task results exhibited even less variability). The figure shows a heat map of the number of times a run was ranked at a given position over all 120 simulation trials---10 trials using the full set of runs to form the pools plus 10 trials for each of 11 teams when omitting a team's uniquely retrieved relevant from the initial pools. The ranks are plotted on the x-axis and the runs on the y-axis where they are sorted by their position in the ranking by the official qrels, using either MAP or Precision@10 as the evaluation measure. The darker a plotted point the more times the run was ranked at that position. The figure makes it clear that a most runs have a single dominant rank. When a run does change ranks, it moves by a modest amount.

\section{Limitations}

% The overall motivation for the TREC Deep Learning Track is to allow benchmarking of new machine learning based retrieval approaches---specifically, deep learning based models~\citep{mitra2018introduction, mitra2021neural}---against traditional IR methods.
The overall motivation for the TREC Deep Learning Track is to allow benchmarking of machine learning based retrieval approaches \citep{Liu:2009,mitra2018introduction,lin2020pretrained} against traditional IR methods, in a setting where large training data is available.
Improving over state-of-the-art on this benchmark may involve designing new neural architectures that, in addition to large scale training data, may also require specialized hardware, \eg, GPUs, to scalably train. Many such deep models that learn from large text data are known to encode problematic societal biases from the corpus or incur significant ecological costs from computationally intensive training~\citep{bender2021dangers}.
Any use of this benchmark for model development should seriously consider such negative externalities and follow strict ethical guidelines.
Alternatively, we also encourage the use of this benchmark for development and study of new methods that may not necessitate large scale training.
The large training data provided may also be used for training models for other tasks.

The queries in this benchmark correspond to real user queries from Bing's search logs.
However, they are restricted to include only English queries that can be answered by a short passage.
This restricts the development of models that may target non-English languages or other user intents---\eg, transactional search.

% Due to resource limitations with regards to judging, the test sets are relatively small (around 50 queries).
% This might limit the sensitivity of this benchmark to smaller improvements.
% It is allowed to combine the two TREC test sets for evaluation to obtain higher statistical power, although the disadvantage would be that any numbers on the combined data set will not be directly comparable to published metrics for baseline methods on the individual test sets.

In spite of these limitations, we believe the TREC DL benchmark is a critical resource for model development and evaluation in the IR community.
%Bias e.g. all English. Memorization. Test sets too small for hypothesis testing / may be insensitive to small improvements?

\section{Conclusion}

We have described the TREC DL resources for document retrieval and passage retrieval, including the training data, Orcas click data, test queries and sample code. Through a case study, we illustrated some best practices for use of the data. Despite using a different labeling scheme, a decision made on the dev set gave us near-optimal performance on both test sets, without making decisions on the test sets which would invalidate the results. Extensions to this approach are to split the dev set into multiple subsets for making different decisions, and running multiple training trials since results of each trial can vary depending on the random seed that is used. Judging on the test sets was quite comprehensive, including the use of HiCAL classifiers to find additional results for judging. Reusability analysis indicated that results are quite stable if we leave out a team, suggesting reasonable performance at evaluating a ranking approach that was not part of the judging procedure. Despite being only one data set, in one language, encouraging work on a particular task, we hope that correct use of the TREC DL sets is useful for future advancements in the field. The TREC Deep Learning Track will continue in 2021.

% \begin{acks}
% \end{acks}

\bibliographystyle{ACM-Reference-Format}
\bibliography{sample-base}

% \appendix









\end{document}